\documentclass[traditabstract, final]{aa}
\usepackage{graphicx}
\usepackage[small]{caption}
\usepackage{epsfig}
\usepackage{natbib}
\usepackage{fancyhdr}
\usepackage{amsmath}
\usepackage{amssymb}
\usepackage{makeidx}
\usepackage{titletoc}
\usepackage[]{subfigure}
\usepackage{lscape}
\usepackage{hyperref}
\usepackage{wasysym}
\usepackage{wrapfig}
\usepackage{longtable}
\usepackage{aas_macros}
\usepackage{txfonts}

\begin{document}

\title{Four and one more: The formation history and total mass of globular clusters in the Fornax dSph}
\titlerunning{The formation history and total mass of globular clusters in the Fornax dSph}

   \author{T.J.L. de Boer\inst{1}
          \and
          M. Fraser\inst{1} 
          }

   \institute{Institute of Astronomy, University of Cambridge, Madingley Road, Cambridge CB3 0HA, UK\\
              \email{tdeboer@ast.cam.ac.uk}
             }

   \date{Received ...; accepted ...}

\abstract{We have determined the detailed star formation history and total mass of the globular clusters in the Fornax dwarf spheroidal using archival HST WFPC2 data. Colour magnitude diagrams are constructed in the F555W and F814W bands and corrected for the effect of Fornax field star contamination, after which we use the routine Talos to derive the quantitative star formation history as a function of age and metallicity.\\
The star formation history of the Fornax globular clusters shows that Fornax~1,~2,~3 and~5 are all dominated by ancient~($>$10 Gyr) populations. Cluster Fornax~1,2 and 3 display metallicities as low as [Fe/H]=$-$2.5 while Fornax 5 is slightly more metal-rich at [Fe/H]=$-$1.8, consistent with resolved and unresolved metallicity tracers. Conversely, Fornax 4 is dominated by a more metal-rich~([Fe/H]=$-$1.2) and younger population at 10 Gyr, inconsistent with the other clusters. A lack of stellar populations overlapping with the main body of Fornax argues against the nucleus cluster scenario for Fornax~4. \\
The combined stellar mass in globular clusters as derived from the SFH is (9.57$\pm$0.93)$\times$10$^{5}$ M$_{\odot}$ which corresponds to 2.5$\pm$0.2 percent of the total stellar mass in Fornax. The mass of the four most metal-poor clusters can be further compared to the metal-poor Fornax field to yield a mass fraction of 19.6$\pm$3.1 percent. Therefore, the SFH results provide separate supporting evidence for the unusually high mass fraction of the GCs compared to the Fornax field population.}


\maketitle

\section{Introduction}
\label{Fnxintroduction}
Dwarf spheroidal galaxies~(dSph) of the Local Group provide an excellent laboratory to test theories of galaxy formation and evolution~\citep[e.g.,][]{Tolstoy09}. The Fornax dSph is no exception, with its unusually complex formation history consisting of stellar populations covering a large range in age and metallicity. The star formation history~(SFH) of the Fornax dSph shows that it has experienced an extended evolution with many periods of active star formation ranging from ancient~(14 Gyr old) to young~(100 Myr)~\citep[e.g.,][]{Gallart052, Coleman08, deBoer2012B,delpino13}. Studies of the kinematics of RGB stars have uncovered the presence of at least three kinematically distinct populations~\citep{Battaglia06, Amorisco12}. Furthermore, spectroscopic studies of Red Giant Branch~(RGB) stars have determined the detailed metallicity distribution function of Fornax, showing a broad distribution with a dominant metal-rich~([Fe/H]$\approx$$-$0.9 dex) component~\citep[e.g.,][]{Pont04, Battaglia06, Letarte10}. \\
Analysis of the spatial distribution shows that Fornax contains a radial population gradient, in which younger, more metal-rich stars are found progressively more toward the centre~\citep{Battaglia06, Coleman08, deBoer2012B}. Furthermore, photometric surveys of Fornax have found several stellar overdensities, some of which have been interpreted as young shell features resulting from the in-fall of a smaller system less than 2 Gyr ago, while another has been confirmed to be a background galaxy cluster \citep{Coleman04,Coleman052,Olszewski06,deBoer2013,Bate15}. Finally, Fornax is one of the few Local Group dSph found to contain a globular cluster~(GC) system~\citep{Shapley382, Hodge61a}.  
\begin{figure}[!b]
\centering
\includegraphics[angle=0, width=0.45\textwidth]{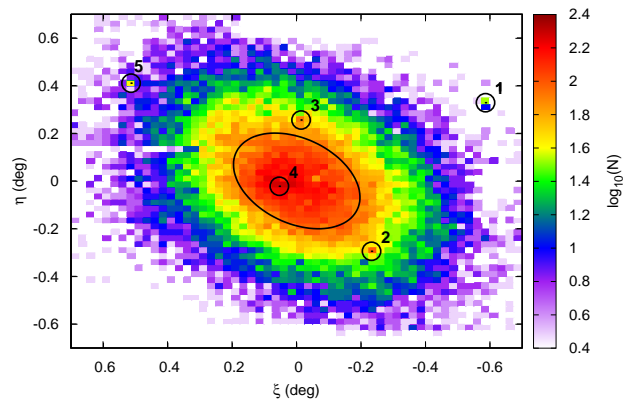}
\caption{Spatial Hess diagram of the RGB population of Fornax from a wide-field dataset~\citep{deBoer2011A}. The circles indicate the position of the GCs in the Fornax dSph, with position according to~\citet{Mackey03b}. The solid ellipse indicates the core radius of the Fornax dSph. \label{spathessRGB}} 
\end{figure}
\\
Models of Fornax have tried to explain the spatial complexity and numerous star formation episodes using tidal encounters with the Milky Way, or dwarf-dwarf merging~\citep[e.g.,][]{Nichols12, Yozin12}. Furthermore, a model involving re-accretion of expelled gas proposed for the Carina dSph could also be invoked to explain the formation of Fornax~\citep{Pasetto11}. However, the exact drivers of the evolution history of Fornax are still unclear, such as the origin of the rapid early enrichment and the occurrence of extended, multiple episodes of star formation. Furthermore, the role of the GC system in the formation of Fornax is also unclear, with some studies concluding that an extremely large fraction~(between 20 and 50 percent, depending in age effects) of the metal-poor stars may be located in the GCs, indicating that the metal-poor Fornax field could be formed solely through the dissipation of clusters during the early formation of the system~\citep{Larsen12b}. By comparing the detailed properties of the GCs to the old stellar populations within Fornax, it will therefore be possible to put greater constraints on the origin of the rapid enrichment and the role played by the GC system. \\
The five GCs of Fornax~(see Figure~\ref{spathessRGB}) have been investigated in great detail, using both resolved and integrated-light techniques~\citep[e.g.,][]{Buonanno85, Dubath92, Buonanno98, Buonanno99, Strader03, Letarte06, Greco07}. Fornax~1,2,3 and 5 have generally been found to be metal-poor, while the innermost GC~(Fornax~4) is more metal-rich. Using accurate high resolution~(HR) spectroscopic observations the mean metallicities of Fornax~1,2,3 and 4 have been determined as~$\langle$[Fe/H]$\rangle$ =$-$2.5, $-$2.1, $-$2.4 and $-$1.6 dex respectively from small samples of stars~\citep{Letarte06, Hendricks15}. Furthermore, \ion{Ca}{ii} triplet spectroscopic metallicities have been obtained for 4 stars in Fornax~1~([Fe/H]=$-$2.81, $-$2.71, $-$2.54 and $-$2.16 dex), 1 star in Fornax~2~([Fe/H]=$-$1.76 dex), 2 stars in Fornax~3~([Fe/H]=$-$2.38 and $-$1.25 dex) and 1 star in Fornax~4~([Fe/H]=$-$0.99 dex), consistent with other metallicity estimates~\citep{Battaglia082}. However, some of these stars are expected to belong to the Fornax field populations instead of the GCs.  Finally, integrated-light studies of the cluster have determined metallicities of  [Fe/H] = $-$2.3, $-$1.4 and $-$2.1 for Fornax 3, 4 and 5 respectively \citep{Larsen12a}. \\
From simple isochrone fitting to resolved Colour-Magnitude Diagrams~(CMDs), the ages of the metal-poor GCs were determined to be consistent with the old population of Fornax~\citep{Buonanno98}. Conversely, the CMD of Fornax~4 is dominated by stars $\approx$3 Gyr younger than the other clusters~\citep{Buonanno99}. Furthermore, estimates of the age from unresolved indices has shown that Fornax~2,~3 and~4 display a similar age, while Fornax~5 is found to be $\approx$2 Gyr younger~\citep{Strader03}. Finally, studies of the mass content of the Fornax GCs has determined that the M/L ratios are comparable to values found in Galactic GCs~\citep{Dubath92, Strader03}. \\
In this paper we use archival Hubble Space Telescope (HST) imaging presented in \citet{Buonanno98,Buonanno99} to determine the detailed SFH of each GC. We will utilise a detailed synthetic CMD method to model the full CMD of each cluster while taking into account the photometric errors and completeness, instead of fitting simple isochrones and ridgelines. This will determine the accurate age and metallicity of each cluster and if they are consistent with a single, short burst of star formation or whether a more complex mix of stellar populations is present. Furthermore, as a byproduct of the SFH determination, we will also be able to determine the total stellar mass of each cluster, which can be compared to the total stellar mass of the Fornax dSph. \\ 
The paper is structured as follows: in section~\ref{data} we outline the data reduction, followed by the sample selection and CMD construction in section~\ref{GCCMD}. Section~\ref{GCdists} discusses the determination of the best-fit distance to each cluster and in section \ref{SFHmethod} we present the best-fit SFH results. Based on the quantitative SFH results, we derive the total stellar masses of the cluster, in relation to the total mass of the Fornax field, presented in section \ref{totmass}. Finally, section~\ref{conclusions} discusses the results obtained from the SFH and the conclusions with regards to the formation and evolution of the Fornax dSph.
\begin{figure*}[!ht]
\centering
\includegraphics[angle=0, width=0.9\textwidth]{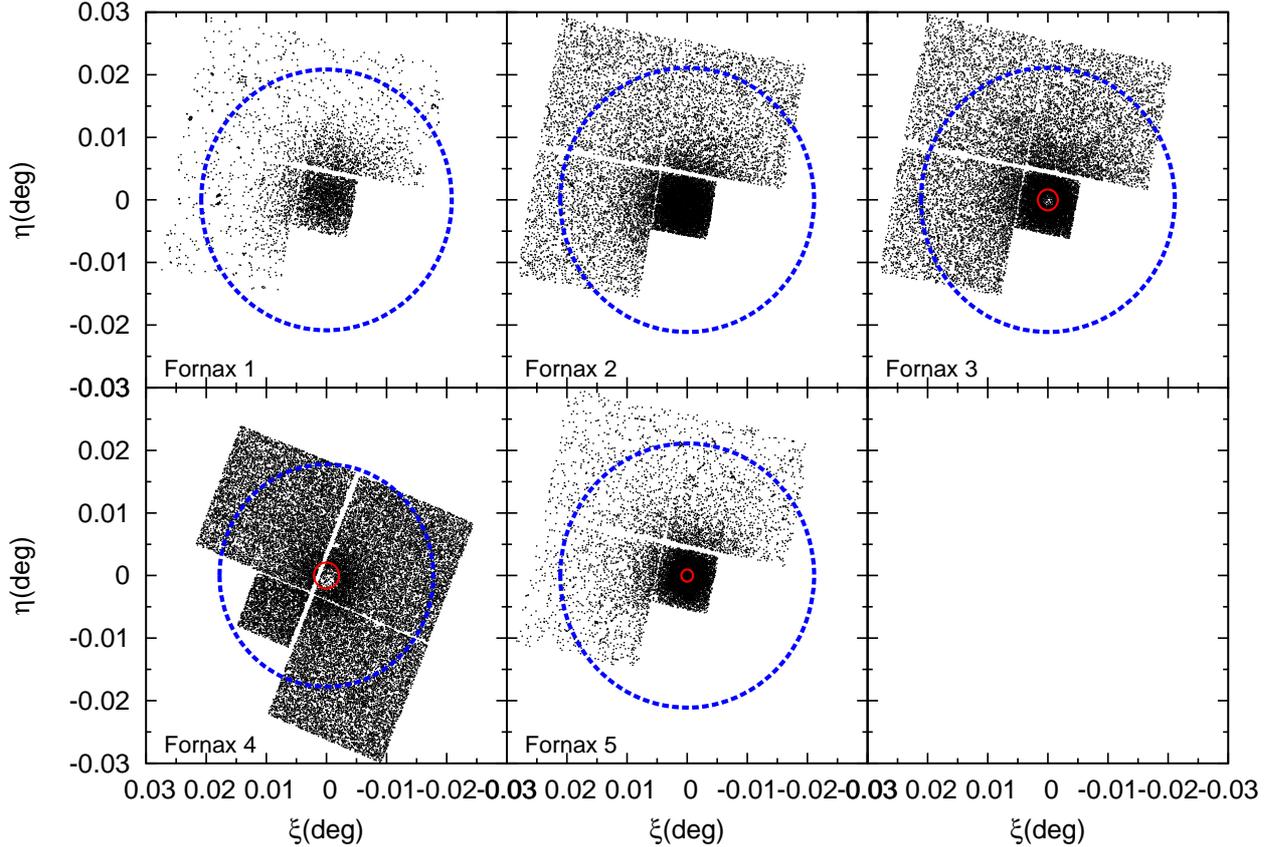}
\caption{Spatial distribution of stars within the HST pointings of each GC, with coordinates centered on the literature positions according to~\citet{Mackey03b}. Blue dashed lines indicate the tidal radius of each cluster while the red solid circles for Fornax 3 and 4 indicate the region heavily affected by stellar crowding. For clusters 1,2,3,5 we limit our sample to the WFPC2 PC chip (smaller square coverage) while for Fornax 4 our sample consists of all stars within a radius of 48$"$~(see section~\ref{GCCMD}). The representative foreground contamination region is constructed from stars outside the blue tidal radius region, except for Fornax 4 where a radius of 48$"$ is adopted . \label{spatialsel}} 
\end{figure*}

\section{Data}
\label{data}
To determine the detailed SFH of the GCs of the Fornax dSph we will make use of archival data. Deep images were obtained with HST/WFPC2 in the broadband F555W and F814W filters for each cluster, as discussed in detail in \citet{Buonanno98,Buonanno99}. The data consists of several long exposures in each filter, accompanied by short exposures to include the otherwise saturated bright stars. For Fornax 1,2,3,5 the cluster centre is placed on the high resolution PC chip, while the centre of Fornax 4 is located mainly on the WF3 chip. \\
All {\it HST}+WFPC2 images were downloaded from the Mikulski Archive for Space Telescopes (MAST\footnote{http://archive.stsci.edu/}), and Point Spread Function (PSF) fitting photometry was performed on them using {\sc hstphot} \citep{Dolphin00}. {\sc hstphot} is a stand-alone photometry package designed for use on WFPC2 data. In each instance, the associated data quality image from MAST was used to mask bad pixels in the image. Cosmic rays and hot pixels were identified and masked, before multiple exposures taken with the same filter and pointing were combined. The offsets between the combined frames were then measured, before {\sc hstphot} was run simultaneously on all images. 
\begin{figure*}[!ht]
\centering
\includegraphics[angle=0, width=0.49\textwidth]{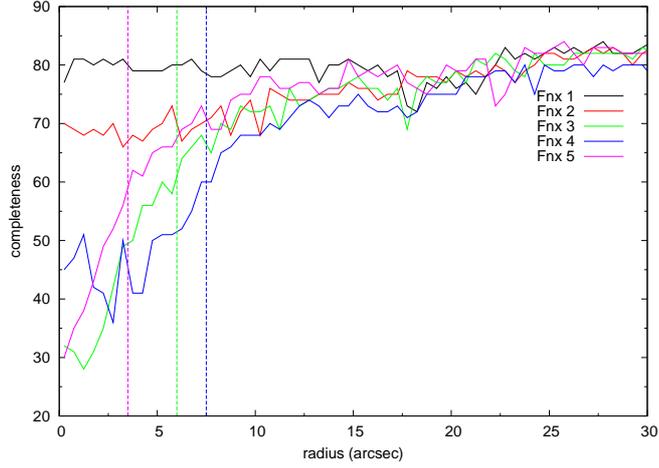}
\caption{The sampling completeness as a function of radius for each GC as derived from artificial star tests. Vertical lines indicate the radial cuts imposed on GC~3,~4 and~5 to restrict the data sample to a completeness greater than $\approx$60 percent. \label{comp_radial}} 
\end{figure*}
\\
A threshold of $3.5\sigma$ in a single frame was used to detect sources, while we required sources to be detected at an overall combined significance of 5$\sigma$ across all images. The recommended parameters were used when running {\sc hstphot} (i.e. refitting the sky background for all images using option 512), with the exception of those taken for Fornax 5, where a local sky background was measured. We also tested the effect of a weighted PSF fit, but found this did not noticeably improve the photometry in the crowded centres of the clusters (as gauged by the width of the main sequence on the resulting CMD). \\
To simulate observational conditions in the synthetic CMDs used to determine the SFH, we make use of artificial star test simulations. To that end, we have carried out a large number of simulations, in which a number of artificial stars~(with known brightness) are placed on the observed images. These images are then analysed in exactly the same way as the original images, after which the artificial stars are recovered from the photometry. In this way we obtain a look-up table, which can be used to accurately model the effects of crowding and completeness as a function of magnitude~\citep[e.g.,][]{Stetson88,Gallart961}. The lookup-table is used in Talos to assign an individual artificial star~(with similar colours and magnitudes) to each star in an ideal synthetic CMD and considering the manner in which this star is recovered to be representative of the effect of the observational biases.
\begin{figure*}[!ht]
\centering
\includegraphics[angle=0, width=0.9\textwidth]{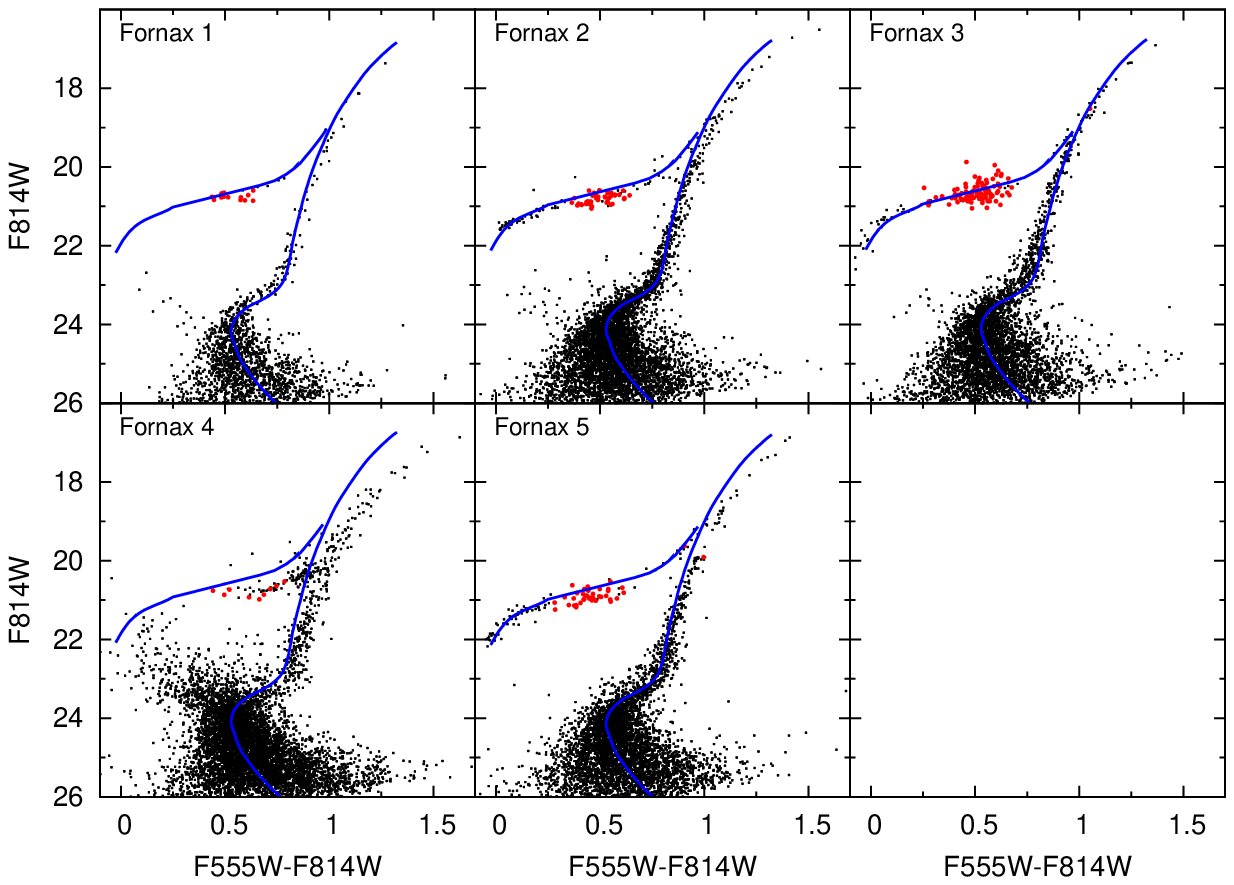}
\caption{Observed CMD diagrams of the five GCs in the Fornax dSph. Red datapoints show confirmed RR Lyrae stars collected from~\citet{Mackey03a,Greco07}. Overlaid isochrones trace a metal-poor ([Fe/H]=$-$2.5 dex) population at a fixed age of 14 Gyr, taken from the Dartmouth Stellar Evolution Database isochrone set~\citep{DartmouthI}. For each isochrone, we adopted a distance as derived from RR Lyrae stars assuming a metallicity as determined from spectroscopic measurements~(see table~\ref{GCpars}). \label{GC_CMD}} 
\end{figure*}

\section{Colour-magnitude diagrams}
\label{GCCMD}
Figure~\ref{spatialsel} presents the spatial distribution of stars within each HST pointings, with coordinates centered on their literature positions according to~\citet{Mackey03b}. The literature tidal radius of each cluster is indicated by the blue circles, making it clear that the HST data does not sample the full stellar distribution of the clusters. Furthermore, several clusters suffer from extensive stellar crowding toward their centres, as shown in Figure~\ref{comp_radial}. The sampling completeness drops significantly for small radii in Fornax 3,4 and 5, possibly impeding a full sampling of the CMD stellar populations in these clusters. Therefore, we excise the overcrowded central regions inside the radius indicated by the vertical lines in Figure~\ref{comp_radial}. To select a stellar sample for which to determine the SFH we limit ourselves to the WFPC2 PC chip (smaller square footprint) for Fornax 1,2,3,5 to avoid introducing inhomogeneous data, and making optimal use of the greater spatial resolution. Finally, we correct the magnitude of each star for the effect of dust extinction using maps of~\citet{Schlegel98} to determine the extinction toward each individual star. \\
Figure~\ref{GC_CMD} shows the resulting F814W, F555W-F814W CMD for each cluster, overlaid with isochrones from the Dartmouth Stellar Evolution Database isochrone set~\citep{DartmouthI} tracing metal-poor~([Fe/H]=$-$2.5 dex) populations at a fixed age of 14 Gyr. For each metallicity, we adopt a value of [$\alpha$/Fe] consistent with the $\alpha$-element distribution determined from high-resolution spectroscopic measurements~\citep{Letarte06,Lemasle14,Hendricks15}. Confirmed RR Lyrae stars for each cluster are labeled with red circles, according to data collected from~\citet{Mackey03a,Greco07}.
\begin{figure*}[!ht]
\centering
\includegraphics[angle=0, width=0.99\textwidth]{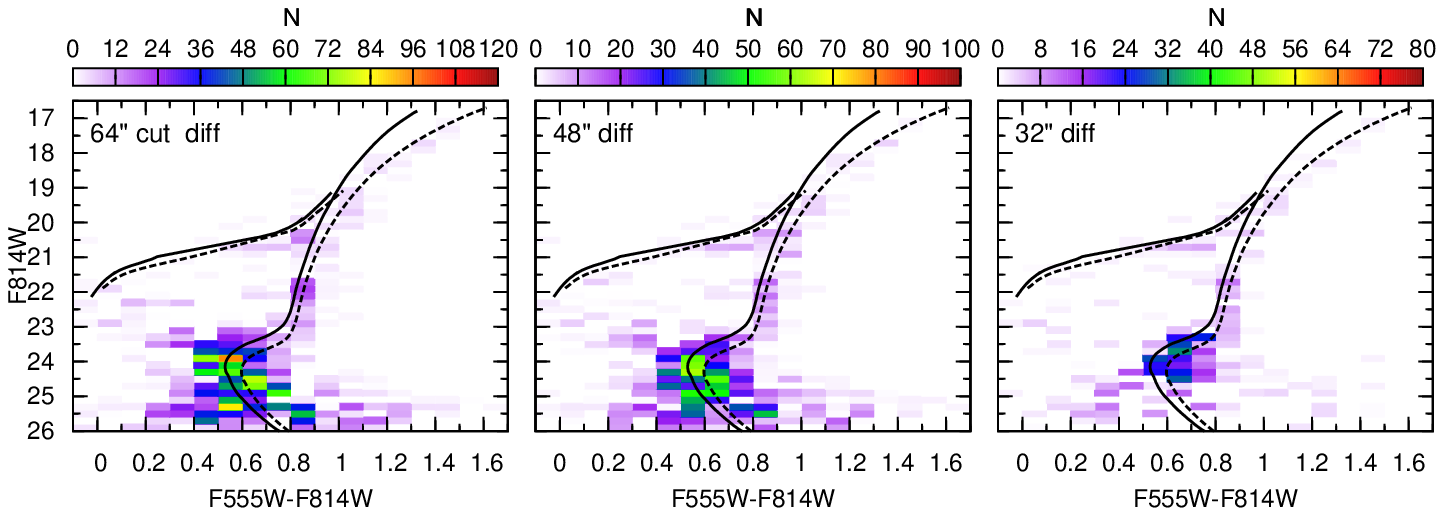}
\caption{Hess diagrams of Fornax 4 after correcting for MW and Fornax field background contamination, with different cuts for the cluster and background population, at 32$"$, 48$"$ and 64$"$ (the tidal radius). A metal-poor ([Fe/H]=$-$2.5 dex, solid) and metal-rich ([Fe/H]=$-$1.5 dex, dashed) isochrone at fixed age of 14 Gyr are overlaid to highlight the shape of population features. \label{GC4rad}} 
\end{figure*}
\\
For each cluster, we also construct Hess diagrams by computing the density of observed stars in small bins in the CMD. The Hess diagrams determined in this way will contain stars of the Fornax GC as well as stars belonging to the MW and the Fornax dSph field population. To correct for the presence of these contaminants, we correct the cluster CMDs by subtracting a representative comparison CMD for each sample. The comparison CMD Hess diagram is scaled using the spatial area subtended on the sky of each cluster and background sample (see Figure~\ref{spatialsel}) before subtraction. For the outer clusters, the coverage of our observations outside the tidal radius (blue circle in Figure~\ref{spatialsel}) is sufficient to sample the background contamination. \\
However, for Fornax~4, the observations cover very little area outside the literature tidal radius, providing a limited sampling of the background CMD features. Due to the central location of Fornax~4, the surface density of contaminants is expected to dominate over the density of cluster stars already well within the tidal radius. Therefore, contamination regions inside the tidal radius are still expected to provide a good background representation, while the larger area covered will lead to a better CMD sampling. To gauge this effect, we constructed BG contamination corrected Hess diagram with three radial data cuts, at 32$"$, 48$"$ and 64$"$ (the tidal radius), shown in Figure~\ref{GC4rad}. For the cut at the tidal radius, the Hess diagram shows stellar populations which are well traced by the overlaid isochrones, but a patchy CMD with strong residuals due to the limited sampling of CMD features in the background region. Conversely, for the innermost cut at half the tidal radius, only limited CMD features are present, indicating that the adopted background region samples and over-subtracts the cluster populations. Therefore, we adopt a radial cut at 48$"$, in which case the Hess diagram recovers a well-defined stellar main sequence turn-off and RGB while minimising the residual Fornax field populations.
\begin{figure*}[!ht]
\centering
\includegraphics[angle=0, width=0.99\textwidth]{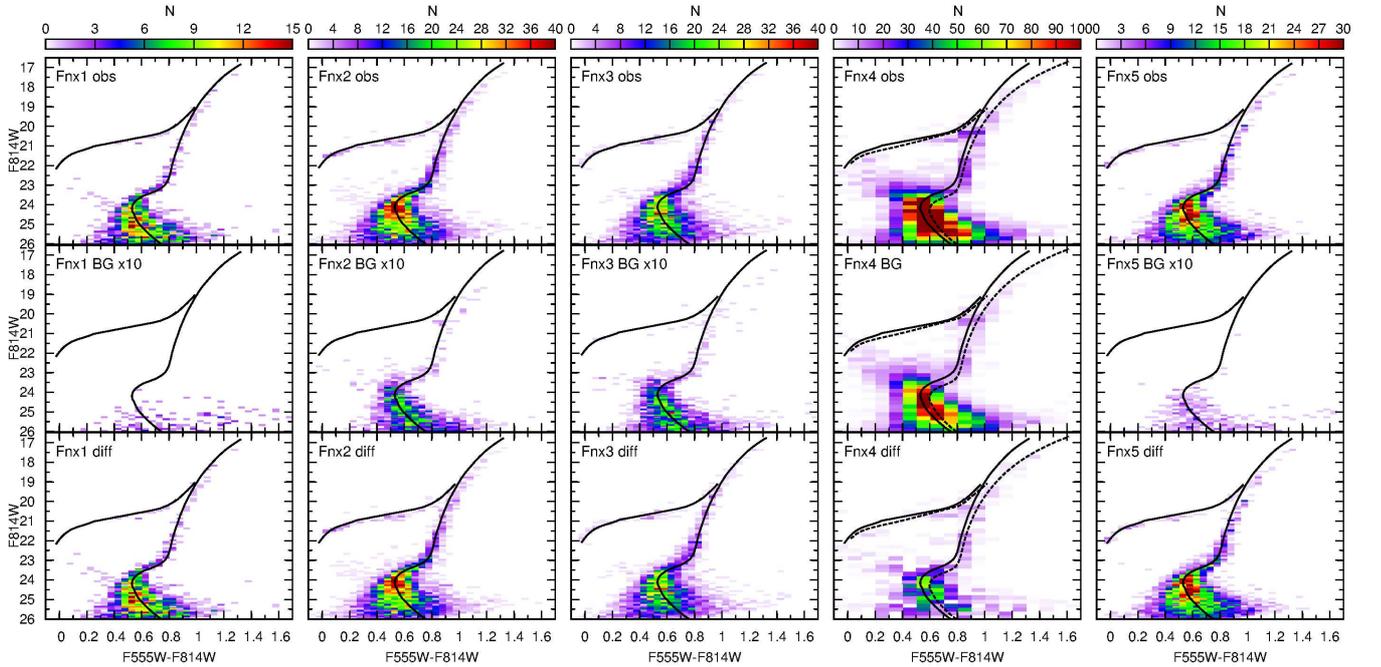}
\caption{Hess diagrams of the GCs in the Fornax dSph, showing the observed data adopted for each cluster~(top row), the data in the MW and Fornax field background reference region (middle row, scaled to the same area as the observed data) and the difference between the two~(bottom row). For the outer four clusters, the level of MW and Fornax field contamination within the cluster aperture is low. Therefore, the reference Hess diagram for these clusters has been multiplied by a factor 10 to highlight features in the CMD. Overlaid (solid) isochrones trace a metal-poor ([Fe/H]=$-$2.5 dex) population at a fixed age of 14 Gyr, taken from the Dartmouth Stellar Evolution Database isochrone set~\citep{DartmouthI}. Furthermore, for Fornax 4 we also show a more metal-rich ([Fe/H]=$-$1.5 dex) isochrone at fixed age of 14 Gyr, as the dashed line. For each isochrone, we adopted the distance given in table~\ref{GCpars}, derived from RR Lyrae stars. \label{GC_hess}} 
\end{figure*}
\\
Figure~\ref{GC_hess} shows the selected Hess diagrams for each cluster, displaying the observed data~(top row), the MW and Fornax field background reference region after scaling to the same area as the observed data (middle row) and the difference between the two~(bottom row). Metal-poor~([Fe/H]=$-$2.5 dex) isochrones at a fixed age of 14 Gyr are overlaid as solid lines, while for Fornax 4 we also show a more metal-rich ([Fe/H]=$-$1.5 dex) isochrone (dashed line). \\ 
The CMDs for Fornax 1,2,3 and 5 displayed in Figures~\ref{GC_CMD} and~\ref{GC_hess} trace a well defined stellar population locus consistent with a low metallicity and old age. Clusters Fornax 1 and 3 appear consistent with the [Fe/H]=$-$2.5 dex isochrone, while the RGB of Fornax~2 and 5 is redder than the isochrone, indicating a slightly higher metallicity (although still quite metal-poor). 
This is in line with results from unresolved metallicity sensitive line indices \citep{Larsen12a}. Clear signs of a blue HB are visible in Fornax 2,3 and 5, indicating the presence of metal-poor stars. No clear HB population can be seen in Fornax~1, which might be due to the relatively low number of stars in this outer GC, leading to a sparsely populated RGB and HB. The four metal-poor GCs display colours at the main sequence turn-off level in agreement with old ages, with metal-poor isochrones tracing the middle of the CMD distribution. The cluster sequences are well-defined in the CMD, with little intrinsic dispersion and no clear signs of multiple composite populations. \\
Finally, the observed CMD of Fornax~4 shows a more complex mix of populations, with a range of populations at the main sequence turn-off level and evidence for low levels of young stars~(see Figure~\ref{GC_CMD}). After subtracting the Hess diagram of the Fornax field star contamination~(see Figure~\ref{GC_hess}), the CMD becomes more defined, with a clear main sequence turn-off consistent with old ages. However, young populations are still visible in Figure~\ref{GC_hess}, which may be residual Fornax field populations due to imperfect background subtraction, given the high levels of contamination. When considering only bright stars, Fornax~4 shows a well-defined thin, red RGB, indicating more metal-rich and potentially younger stellar populations. No clear indication of a metal-poor RGB is seen in the CMD, suggesting that this cluster is different from the other 4 Fornax GCs.

\section{Distance}
\label{GCdists}
A crucial parameter for the determination of the detailed SFH is the distance to each individual GC. An incorrect distance would lead to a systematic offset between the model and observed CMD, which leads to a bias in the stellar populations of the cluster by for instance adopting too young and too metal-poor populations in the case of a too large distance. The Fornax dSph is located at a distance of 147$\pm$4 kpc~((m-M)$_{\mathrm{V}}$=20.84$\pm$0.04), determined using RR Lyrae stars, in good agreement with other measurements using the infrared tip of the Red Giant Branch~(RGB) method and the Horizontal Branch~(HB) level~\citep{Greco05,Rizzi07,Pietrzynski09}. The GCs of Fornax have also been the subject of variability studies, leading to the detection of RR Lyrae stars in each cluster, as plotted in Figure~\ref{GC_CMD}~\citep{Mackey03a,Greco07}. \\
However, the distance toward each cluster is not only dependent on the mean RR Lyrae brightness but also on the metallicity of the cluster. The size of this effect can be significant, with a change in distance modulus of $\approx$0.2 mag when switching from a metallicity of [Fe/H]=$-$2.5 dex to [Fe/H]=$-$1.5 dex.  Therefore, assuming an incorrect metallicity for the clusters would lead to a bias in the distance and therefore a bias in the SFH results. \\
To determine the distance toward each cluster, we make use of the spectroscopic metallicity studies conducted for each cluster. For Fornax 1,2 and 3 we adopt metallicities from high-resolution spectroscopy of individual stars from~\citet{Letarte06} and for Fornax 4 and 5 we adopt the most recent integrated-light metallicity estimates from~\citet{Larsen12a}. Subsequently, we determine the intrinsic brightness of RR Lyrae stars of each metallicity using relations by \citet[e.g.][]{Cacciari03,Clementini03,Gratton04}. Following this, comparison between the observed mean brightness of the RR Lyrae stars as presented in~\citet{Mackey03a,Greco07}, and taking into account the average visual dust extinction uniquely determines the distance towards each cluster. The adopted parameters for the Fornax GCs and their uncertainties are presented in table~\ref{GCpars}. The table shows that all clusters are consistent within the uncertainties with a location inside the main body of Fornax. The outermost cluster Fornax~1 is located at a distance consistent or slightly behind the Fornax centre, whereas the other four clusters are more consistent with being seen slightly in front of the main body. 
\begin{table}[!ht]
\caption[]{Adopted parameters for each globular cluster, including the mean visual magnitude for RR Lyrae stars~\citep{Mackey03a,Greco07}, average visual dust extinction~\citet{Schlegel98}, spectroscopic metallicities~\citep{Letarte06, Larsen12a} and derived distances. Uncertainties on the adopted distances have been computed taking into account the uncertainties on the variables as listed here. We also give the fraction of the total stellar mass sampled by the observations. \label{GCpars}}
\begin{center}
\resizebox{0.5\textwidth}{!}{
\begin{tabular}{ccccccc}
\hline\hline
GC & $\left<{V_{RR}}\right>$ & A$_{V}$ & [Fe/H] & distance {\tiny(kpc)} & m-M & f$_{sampling}$ \\
\hline
1 & 21.27$\pm$0.01 & 0.058$\pm$0.004 & -2.5$\pm$0.1 & 147.2$\pm$4.1 & 20.84$\pm$0.06 & 0.58 \\
2 & 21.34$\pm$0.01 & 0.102$\pm$0.005 & -2.1$\pm$0.1 & 143.2$\pm$3.3 & 20.78$\pm$0.05 & 0.68 \\
3 & 21.24$\pm$0.01 & 0.081$\pm$0.003 & -2.4$\pm$0.1 & 141.9$\pm$3.9 & 20.76$\pm$0.06 & 0.41 \\
4 & 21.43$\pm$0.03 & 0.075$\pm$0.005 & -1.4$\pm$0.1 & 140.6$\pm$3.2 & 20.74$\pm$0.05 & 0.37 \\
5 & 21.33$\pm$0.01 & 0.070$\pm$0.001 & -2.1$\pm$0.1 & 144.5$\pm$3.3 & 20.80$\pm$0.05 & 0.57 \\
\hline 
\end{tabular}
}
\end{center}
\end{table}

\section{Star formation history}
\label{SFHmethod}
\begin{figure}[!ht]
\centering
\includegraphics[angle=0, width=0.4\textwidth]{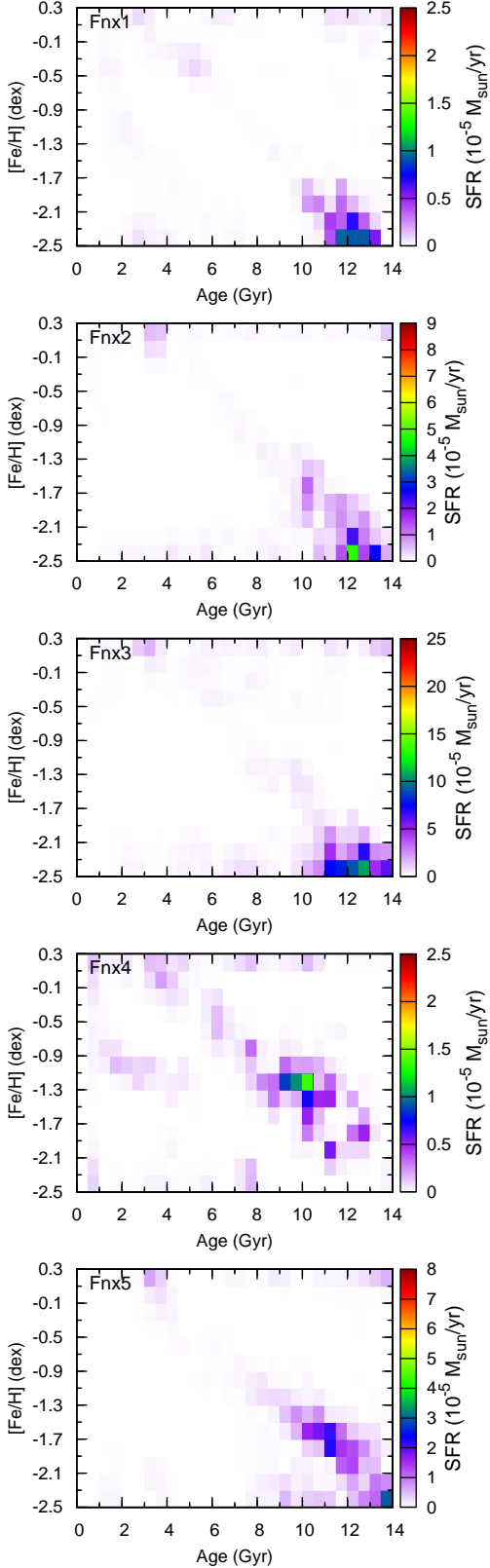}
\caption{The full 2D SFH solution derived for each of the 5 Fornax GCs, from fits to the CMDs as shown in Figure~\ref{GC_hess}. Values for the SFR have been corrected for the missing mass fraction using values given in table~\ref{GCpars}. \label{FnxSFRGC}} 
\end{figure}

\begin{figure*}[!thb]
\centering
\includegraphics[angle=0, width=0.95\textwidth]{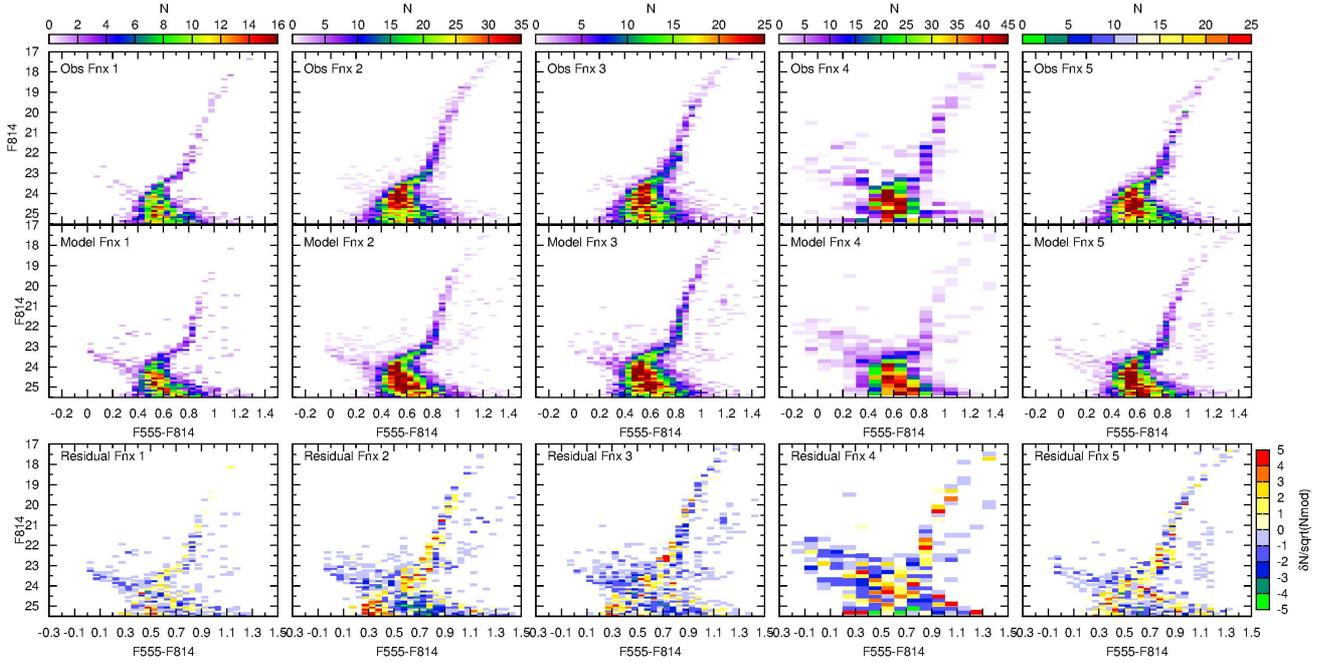}
\caption{{\bf Top:} the observed, foreground corrected Hess diagrams as used in the SFH determination for each cluster. {Middle:} the best-fit model CMD obtained by Talos. {\bf Bottom:} the difference between the observed and best-fit CMD, expressed as a function of the uncertainty in each CMD bin. \label{residuals}} 
\end{figure*}

The SFH of each GC will be determined using the SFH fitting code Talos, described in detail in~\citet{deBoer2012A}. The SFH is derived on the basis of the extinction-free (F814W, F555W$-$F814W) CMD, corrected for the presence of foreground MW and Fornax field star contamination. A uniform bin size of 0.05 in colour and 0.1 in magnitude is used for all clusters except Fornax~4. For this cluster, we double the bin sizes due to the sparse sampling of CMD features after correcting for Fornax field contamination. No spectroscopic metallicity distribution function is used in the SFH determination, due to the limited sample of spectroscopic observations available. We adopt the Dartmouth Stellar Evolution Database isochrone set~\citep{DartmouthI} in the SFH fitting, since this set was also used in the derivation of the SFH of the Fornax field population~\citep{deBoer2012B}. For the SFH solution, metallicities are allowed to range from $-$2.5$\le$[Fe/H]$\le$+0.3 dex with a spacing of 0.2 dex, for ages between 0.5 and 14 Gyr with a spacing of 0.5 Gyr. While it is possible for a cluster to have a metallicity below [Fe/H]=$-$2.5 dex, isochrones which such low metallicities become degenerate, producing indistinguishable CMDs. Therefore, we only consider metallicities down to this limit. \\
The fitting procedure results in a 3D solution giving the star formation rate~(SFR) for each age and metallicity, which is present for each cluster in Figure~\ref{FnxSFRGC}. By projecting this solution onto the age axis we obtain the SFH, while projection onto the metallicity axis gives the chemical evolution history, presented in Figure~\ref{FnxSFHGC}. To assess the width of SFR peaks in age and metallicity space, we fit Gaussian profiles to the SFH to determine the mean position of the central peak as well as the variance. The Gaussian profiles are also shown as dashed lines in Figure~\ref{FnxSFHGC}, with fitted values for age and metallicity presented in table~\ref{Gausspars}.
\\
Following the fitting procedure, Figure~\ref{residuals} shows the observed and synthetic (F814W, F555W$-$F814W) CMD of each Fornax cluster and the residual in each bin in terms of Poissonian uncertainties. The residuals show that the synthetic CMDs are largely consistent with the observed CMDs within 3 sigma in each bin barring a small fraction of bins with positive residuals, indicating an overall good fit of the data. The synthetic CMDs correctly reproduce all evolutionary features in the lower CMD, including the location and spread of the main sequence turn-off, subgiant branch and the slope of the RGB. \\
Figures~\ref{FnxSFRGC} and~\ref{FnxSFHGC} show that Fornax 1,2,3 and 5 are all dominated by ancient, metal-poor populations, consistent with the CMDs presented in Figure~\ref{GC_hess}, while Fornax~4 display a peak at more intermediate age and metallicity. Each cluster displays a somewhat extended sequence in age-metallicity space, due to the ever-present age-metallicity degeneracy which results in different combinations of age and metallicity producing a very similar CMD. This degeneracy is not fully broken when sampling the full CMD at high resolution, especially at old ages. Nonetheless, the combination of a well-sampled narrow main sequence turn-off and thin RGB places strong constraints on the possible age and metallicity of the cluster, as evidenced by the peaks in star formation rate~(SFR). The SFH of each cluster also displays a small burst of star formation at young~(2$-$4 Gyr) ages, which is due to the presence of small numbers of blue straggler stars. These small peaks will be ignored during any of the following discussion. \\
The SFH of Fornax~1 peaks at an age of $\approx$12 Gyr, favouring an age slightly younger than the maximum allowed. The chemical evolution history peaks toward the lower limit in our parameter space, indicating this is indeed a metal-poor, ancient GC, consistent with a scenario in which it formed in a single, short episode of star formation. The SFH of Fornax~3 is very similar, although the age distribution is broader and can be linked to the relatively wide RGB observed in Figure~\ref{GC_hess}. Likewise, Fornax~2 is also an ancient cluster with an age of $\approx$12 Gyr and a metallicity that peaks at the limit of our parameter space. This is slightly more metal-poor than the results from spectroscopic measurements, although broadly consistent within the uncertainties of the peak. \\
Fornax~5 peaks at a slightly younger age than the other metal-poor clusters, but shows a distribution in age wider than any of the other old clusters. This wide distribution might result in a younger average, light-weighted age as observed by \citet{Strader03}. The chemical evolution history of Fornax~5 is wide, with a high SFR at $-$2.5 dex and a more defined Gaussian peak at [Fe/H]=$-$1.7 dex, consistent with measurements using unresolved metallicity indicators~\citep{Larsen12a}. Finally, the SFH of Fornax~4 displays a large number of patchy low SFR populations, likely a result of residual contamination due to the underlying Fornax field, as seen in the CMD in Figures~\ref{GC_CMD} and~\ref{GC_hess}. Nonetheless, the chemical evolution history shows a well defined metallicity distribution peaking at [Fe/H]=$-$1.2 dex and a clear peak in the SFH at an age of $\approx$10 Gyr~(see table~\ref{Gausspars}). These parameters are consistent with both the red RGB and relatively blue MSTO in Figure~\ref{GC4rad}. 
\begin{table}[!ht]
\caption[]{Best-fit mean location and standard deviation for age and metallicity parameters according to Gaussian profile fits to the SFH shown in Figure~\ref{FnxSFHGC}. The best-fit parameters obtained for the recovered SFH for mock stellar populations are also shown. \label{Gausspars}}
\begin{center}
\resizebox{0.5\textwidth}{!}{
\begin{tabular}{ccccc}
\hline\hline
GC & age$_{SFH}$ & [Fe/H]$_{SFH}$ & age$_{mock}$ & [Fe/H]$_{mock}$  \\
\hline
1 & 12.1$\pm$0.8 & $-$2.5$\pm$0.3 & 11.9$\pm$0.6 & $-$2.5$\pm$0.3  \\
2 & 12.2$\pm$1.0 & $-$2.5$\pm$0.3 & 11.9$\pm$0.8 & $-$2.5$\pm$0.3  \\
3 & 12.3$\pm$1.4 & $-$2.5$\pm$0.2 & 12.0$\pm$0.9 & $-$2.5$\pm$0.3  \\
4 & 10.2$\pm$1.2 & $-$1.2$\pm$0.2 & 10.3$\pm$0.7 & $-$1.2$\pm$0.1  \\
5 & 11.5$\pm$1.5 & $-$1.7$\pm$0.3 & 11.9$\pm$1.1 & $-$1.8$\pm$0.2  \\
\hline 
\end{tabular}
}
\end{center}
\end{table}

\section{Testing for extended populations}
\label{mocktests}
Having determined the detailed SFH of the Fornax GCs, we consider the question whether the clusters show any signs of composite multiple populations. An important consideration when judging the population spread in the clusters is the intrinsic uncertainty on age and metallicity due to the adopted SFH method. The ability to resolve an episode of star formation depends mainly on the photometric depth and associated photometric errors of the observed data. Furthermore, the method of determining the uncertainties of the solution also results in a smoothing of the SFH, which gives limits to the resolution of the SFH. To determine the intrinsic resolution we construct a set of mock synthetic CMDs for single burst populations, similar to the scheme proposed by~\citet[e.g.,][]{Hidalgo11}. The manner in which these single burst populations are recovered will determine the intrinsic resolution of the SFH. 
\begin{figure*}[!thb]
\centering
\includegraphics[angle=0, width=0.75\textwidth]{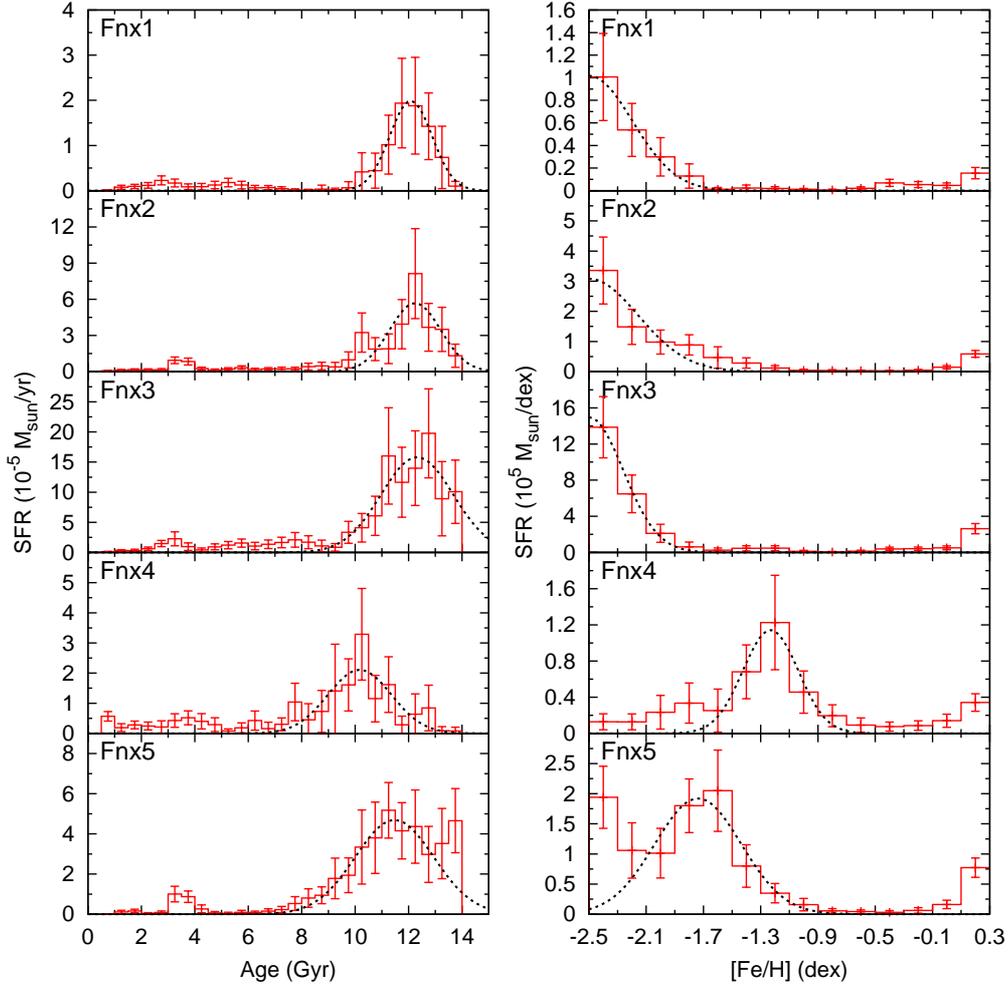}
\caption{The SFH~(left) and chemical evolution history~(right) of the 5 GCs of the Fornax dSph, as obtained from the HST F814W, F555W$-$F814W CMD. Values for the SFR have been corrected for the missing mass fraction using values given in table~\ref{Gausspars}. Dashed lines show the best-fitting Gaussian profile to the age and metallicity distribution, with parameters given in table~\ref{Gausspars}. \label{FnxSFHGC}} 
\end{figure*}
\\
We construct a synthetic CMD for each cluster, with an age and metallicity corresponding to the middle of the population bin in which the SFH peak is located, as shown in Figures~\ref{FnxSFHGC} and listed in table~\ref{Gausspars}. Subsequently, we convolve the CMD with artificial star test completeness results and photometric error curves from the observed CMD, to simulate the observational conditions of each Fornax GC. The synthetic CMDs are subsequently analysed in exactly the same manner as the data, and the manner in which the single bursts are recovered indicates the intrinsic uncertainty due to the adopted method. \\
The recovered SFH for the synthetic bursts is shown in Figure~\ref{GCresolution}, for each cluster, along with the SFH corresponding to a perfect recovery for comparison as the solid black histogram. The recovered SFH is fit by a Gaussian distribution, from which we determine the age of the central peak as well as the variance, as shown in table~\ref{Gausspars}. \\
Figure~\ref{GCresolution} shows that the central peak is correctly recovered for each cluster, barring small offsets toward younger ages for the most metal-poor clusters, likely due to edge effects in the metallicity distribution. The Gaussian fits show that the widths of the distribution of recovered mock populations are similar to those of the cluster SFHs in Figure~\ref{FnxSFHGC}. Therefore, we find no clear signs of extended populations in any of the clusters, given the smoothing induced by the data quality and SFH method. Age distributions in the recovered SFHs are slightly narrowed than those in cluster SFHs, which might indicate that the clusters are more extended in age than delta function single burst populations. However, the width of the distributions might also be affected by residual contamination, the presence of blue straggler stars and unresolved binaries.

\section{Total mass}
\label{totmass}
\begin{figure}[!ht]
\centering
\includegraphics[angle=0, width=0.49\textwidth]{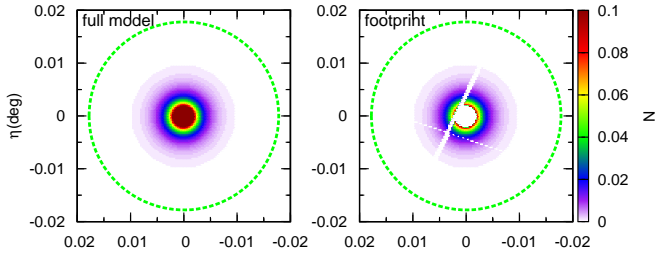}
\caption{{\bf Left:} Normalised spatial distribution for cluster Fornax~4 with a stellar density according to King profile fits from~\citet{Mackey03b}. The green (dashed) circle indicates the tidal radius. {\bf Right:} Spatial distribution for the cluster convolved with the actual coverage of the HST pointing, and the rejection of a central region due to incomplete data. Comparison between the two distributions allows us to determine the fraction of stars (and therefore mass) missing from the adopted sample. \label{GC4_Kingmodel}} 
\end{figure}

Using the results of the SFH, it is possible to derive the total mass formed in each GC, directly from the CMD fitting. By integrating the SFR of each stellar population, we obtain the total mass in stars formed over the mass range 0.1$-$120 M$_{\odot}$, under the assumption of a Kroupa IMF~\citep{Kroupa01}. However, to obtain the full stellar mass of the GCs we need to account for the fact that we do not sample the full stellar distribution of each cluster. To that end, we turn to the stellar density models derived for the cluster by \citet{Mackey03b}, where King profiles were fit to the stellar number counts as a function of radius. \\
We can use these best-fit models to compute the fraction of the stellar mass sampled by our target selection and SFHs. For each cluster, we construct the 2D spatial distribution of stars according to the King profile parameters of \citet{Mackey03b}. Subsequently, we convolve the distribution with the HST footprint as well as the radial cuts applied to remove spatial regions heavily influenced by stellar crowding. The observed and model distributions obtained in this way are shown in Figure~\ref{GC4_Kingmodel} for Fornax~4 to illustrate the process. We use the comparison between the total number of stars in each distribution to determine the fraction of stars and therefore mass missing from the SFH of each cluster, shown in table~\ref{GCpars}. In the case of mass segregation with the Fornax clusters, one can wonder if the fits derived using relatively bright and massive stars provide a valid fit to the spatial distribution of stars of all masses. However, to probe this effect, better spatial resolution data extending down to the lower mains sequence is needed.
\begin{figure*}[!thb]
\centering
\includegraphics[angle=0, width=0.75\textwidth]{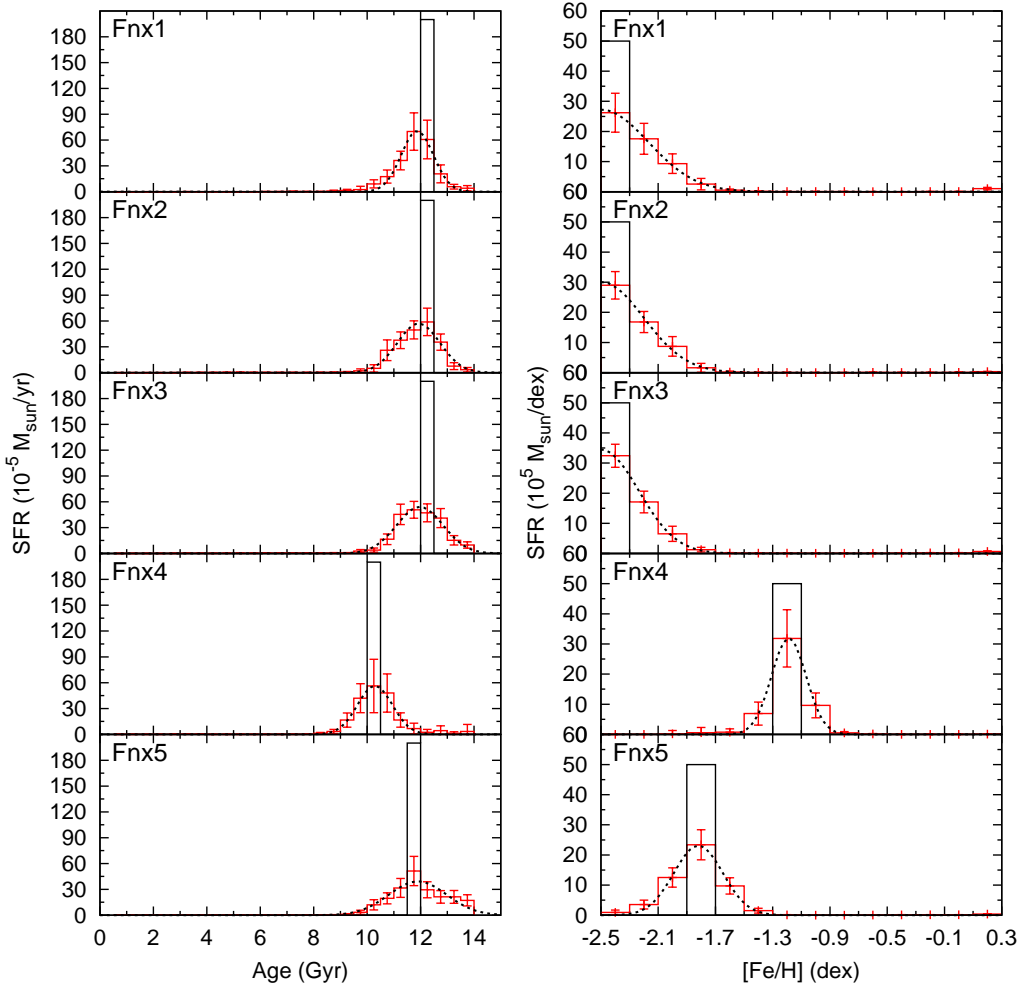}
\caption{Input and recovered SFH~(left) and chemical evolution history~(right) of mock stellar populations for each cluster, based on the peaks of star formation in Figure~\ref{FnxSFHGC}. The black, solid histogram shows the input SFH, given the adopted binning of the solution. The red histograms show the recovered SFH, while the black dashed lines show the best-fit Gaussian distribution, with mean and variance listed in table~\ref{Gausspars}. \label{GCresolution}} 
\end{figure*}
\\
The resulting total mass derived for each GC after correcting for the missing mass fraction is shown in table~\ref{GCmasses}. The computed total stellar masses show that Fornax~3 is the most massive cluster in the Fornax dSph, followed by Fornax~5 and 2. Unsurprisingly, the outermost cluster Fornax~1 has the lowest mass of the GC system, resulting in a sparsely populated CMD in Figure~\ref{GC_hess}. The computed masses are in reasonable agreement with the results of \citet{Mackey03b} given the errorbars, and the order of cluster masses is also in line with unresolved absolute magnitudes \citep{Larsen12b}. \\
Following the work of~\citet{Larsen12b} it is useful to compare the masses of the Fornax GCs to the total stellar mass in the Fornax dSph field. To compute the total stellar mass in Fornax we turn to the data presented in~\citet{deBoer2012B}, where SFHs were determined within the same framework as the results presented here. Given the old age of stellar populations found in the outskirts of Fornax, we re-determine the SFH using the deep V,B$-$V data, to gain an optimum sampling of the oldest main sequence turn-off region. The SFH obtained in this way can be integrated to obtain the total stellar mass out to an elliptical radius of 0.8 degrees, which falls short of the tidal radius of 1.18 degrees \citep{Irwin95}. To obtain the total stellar mass out to the tidal radius, we apply a correction to the stellar mass obtained from the SFH by assuming that the outer regions of Fornax display the same population mix and density as the outermost ellipse in~\citet{deBoer2012B}. The total stellar mass in the outermost region (as derived from the SFH) is scaled to the spatial area of the missing part and added on top of the derived SFH mass to obtain a total stellar mass within the tidal radius of (3.82$\pm$0.12)$\times$10$^{7}$ M$_{\odot}$. This value is lower than the results computed in~\citet{deBoer2012B} as a result of the depth of the B$-$V data, leading to better constraints on the oldest populations, making up the bulk of the mass in Fornax. \\
The combined mass in the Fornax GCs is (9.57$\pm$0.93)$\times$10$^{5}$ M$_{\odot}$ which corresponds to 2.5$\pm$0..2 percent of the total stellar mass in Fornax. However, the bulk of the stars in Fornax is metal-rich~([Fe/H]$>-$1.5 dex) while the majority of the clusters are metal-poor. Therefore, it is useful to compare the total mass of the metal-poor clusters of (8.81$\pm$0.92)$\times$10$^{5}$ M$_{\odot}$ to the SFH-derived metal-poor~([Fe/H]$<$$-$2 dex) stellar mass in the Fornax main body of (44.9$\pm$5.3$)\times$10$^{5}$ M$_{\odot}$ to give a mass fraction of 19.6$\pm$3.1 percent. Therefore, the metal-poor globular clusters encompass a substantial fraction of the stellar mass of the Fornax dwarf galaxy. This recovered value for the total stellar mass of the cluster is in agreement with the results of~\citet{Larsen12b} when corrected for age effects. 

\section{Discussions and conclusions}
\label{conclusions}
The SFH derived for the Fornax GCs shows that Fornax~1,2,3 and 5 are all old and metal-poor clusters, while Fornax~4 is dominated by more metal-rich and younger stars. The metallicities derived for the GCs~(Figure~\ref{FnxSFHGC}) are in agreement with integrated-light and HR spectroscopic measurements~\citep[e.g.,][]{Letarte06, Larsen12a, Hendricks15}. However, the results obtained from the SFH fitting further quantify the age and metallicity distribution of the GCs and determine the mass in stars formed at each epoch. \\
The SFH of Fornax~1,2 and 3 display a clear peak at [Fe/H]$\approx$$-$2.5 dex, with ages consistent with an ancient~($\ge$10 Gyr) population. Therefore, results obtained from the CMD confirm their status as some of the most metal-poor GCs known~\citep{Letarte06}. Results for Fornax~3 indicate that it is the most massive cluster in the Fornax dSph, with a mass more than twice that of any of the other clusters. The larger mass of Fornax~3 has not resulted in faster enrichment and a consequently higher metallicity. Instead, the SFH results are in good agreement with the lower mass clusters Fornax~1 and~2. \\
The SFH of Fornax~4 is found to be different from that of the other clusters in Fornax, with a clear peak at an intermediate age of 10 Gyr and a metallicity of $-$1.2 dex. Therefore, this cluster is substantially more metal-rich than the other clusters, in agreement with previous results based on integrated-light spectroscopy. Previous isochrone analysis of the age of Fornax~4 determined it to have an age $\approx$3 Gyr younger than the other Fornax GCs, which roughly agrees with the age peak in our SFH~\citep{Buonanno99,Hendricks15}.
\begin{table}[!ht]
\caption[]{Total mass in stars formed in each GC over the mass range 0.1$-$120 M$_{\odot}$. The mass of stars formed in the metal-poor component~([Fe/H]$<-$2.0) contribution is also given. \label{GCmasses}}
\begin{center}
\begin{tabular}{cc}
\hline\hline
GC & Total stellar mass \\
       &  10$^{5}$ M$_{\odot}$ \\
\hline
Fnx~1 & 0.42$\pm$0.10  \\
Fnx~2 & 1.54$\pm$0.28  \\
Fnx~3 & 4.98$\pm$0.84  \\
Fnx~4 & 0.76$\pm$0.15  \\
Fnx~5 & 1.86$\pm$0.24  \\
\hline 
\end{tabular}
\end{center}
\end{table}
\\
The location close to the centre of Fornax has given rise to multiple explanations on the nature of Fornax~4. In particular,~\citet{Hardy02} and \citet{Strader03} argue that it is the nucleus of the Fornax dSph, explaining its unusual populations, high metallicity and central location. The spatial Hess diagram of Fornax in Figure~\ref{spathessRGB} shows that the position of Fornax~4 coincides with the highest density region of  Fornax, which is also further associated with the youngest populations of Fornax distributed in a shape inconsistent with that of the old populations \citep{Battaglia06}. Our SFH shows that Fornax~4 is dominated by a single stellar population at an age of $\approx$10 Gyr with little dispersion in age and metallicity. However, there is residual star formation at other ages, due to the high levels of Fornax field contamination. Nonetheless, the lack of ancient, metal-poor populations argues against the nucleus scenario. This is consistent with the radial velocity of the Fornax~4, which is distinctly different from that of the main body of Fornax \citep{Larsen12a, Hendricks15}. If this cluster is indeed a genuine globular cluster seen in projection to the centre of Fornax, it is still unclear why the cluster shows populations so different from others GCs in Fornax. A more complete photometric mapping of the inner regions of Fornax~4 with higher spatial resolution is needed to unambiguously separate the cluster population from the field contamination, preferably supported by spectroscopic observations of the bright cluster giants. 
\\
Finally, the CMD of Fornax~5~(see Figure~\ref{GC_hess}) shows a relatively red RGB that can be explained by a slightly higher metallicity and a wider age distribution than the other clusters, in line with light-weighted ages by~\citet{Strader03}. We derive a relatively low mass with regards to integrated-light observations, which may be a result of the M/L ratio assumed for the cluster~\citep[e.g.,][]{Mackey03b}. A younger age for the cluster would result in a smaller M/L ratio and a lower total stellar mass. \\
The total stellar masses of the Fornax GCs computed from the SFH is (9.57$\pm$0.93)$\times$10$^{5}$ M$_{\odot}$ which corresponds to a small fraction of the total stellar mass in the Fornax dSph. Comparison between the metal-poor clusters and the metal-poor field gives a mass fraction of 19.6$\pm$3.1 percent, providing separate supporting evidence for the results of~\citet{Larsen12b}. This unusually high mass fraction of the Fornax GCs compared to the field can be used to put valuable constraints on self-enrichment scenarios in which GCs were initially much more massive and lost most of their stars to the Fornax field \citep[e.g.][]{Schaerer11,Bekki11,Larsen12b}. Clearly, unlocking the full formation history of the Fornax dSph and its individual components will provide vital insights into the complex formation scenario of the high end of the dwarf galaxies mass scale.

\section{Acknowledgements}
\label{acknowledgements}
The research leading to these results has received funding from the European Research Council under the European UnionÕs Seventh Framework Programme (FP/2007-2013) / ERC Grant Agreement n. 308024. This work was partly supported by the European Union FP7 programme through ERC grant number 320360. The authors would like to thank the anonymous referee for his/her comments, that greatly helped to improve the paper.

\bibliographystyle{aa}
\bibliography{../Bibliography}

\end{document}